\documentstyle[12pt,psfig]{article}
\newcommand{\EQ}{\begin{equation}}
\newcommand{\EN}{\end{equation}}
\newcommand{\bea}{\begin{eqnarray}}
\newcommand{\eea}{\end{eqnarray}}

\newcommand{\th}{\theta}
\newcommand{\var}{\varepsilon}

\newcommand{\lab}{\label}

\begin{document}
\topmargin 0pt
\oddsidemargin 5mm
\renewcommand{\thefootnote}{\arabic{footnote}}
\newpage
\setcounter{page}{0}
\begin{titlepage}
\vspace{0.5cm}
\begin{center}
{\large {\bf The Field Theory of the $q\to4^{+}$ Potts Model}}\\
\vspace{1.8cm}
{\large G. Delfino$^a$ and John Cardy$^{b,c}$} \\ \vspace{0.5cm}
{\em $^a$ Laboratoire de Physique Th\'eorique et Hautes Energies}\\
{\em Universit\'e Pierre et Marie Curie, Tour 16 $1^{er}$ \'etage, 4 place 
Jussieu}\\
{\em 75252 Paris cedex 05, France}\\
{\em $^b$ Theoretical Physics, University of Oxford}\\
{\em 1 Keble Road, Oxford OX1 3NP, United Kingdom}\\
{\em $^c$ All Souls College, Oxford} 
\end{center}
\vspace{1.2cm}

\renewcommand{\thefootnote}{\arabic{footnote}}
\setcounter{footnote}{0}

\begin{abstract}
\noindent
The $q$-state Potts model in two dimensions exhibits a first-order
transition for $q>4$. As $q\to4^{+}$ the correlation length at this transition
diverges. We argue that this limit defines a massive integrable quantum field
theory whose lowest excitations are kinks connecting $4+1$ degenerate
ground states. We construct the $S$-matrix of this theory and the
two-particle form factors, and hence estimate a number of universal
amplitude ratios. These are in very good agreement with the results of
extrapolated series in $q^{-1/2}$ as well as Monte Carlo results for $q=5$.

\end{abstract}

\vspace{.3cm}

\end{titlepage}

\newpage
The $q$-state Potts model, defined by the lattice Hamiltonian
\cite{Potts}
\EQ
H=-J\sum_{\langle x,y\rangle}\delta_{s(x),s(y)}\,\,,
\nonumber
\EN
where the spin variable $s(x)$ assumes $q$ different values (colours), 
continues to be of fundamental importance in the description of a 
large variety of critical phenomena, ranging from ferromagnetism to percolation
and adsorbed monolayers \cite{Wu}. It was shown by Baxter that in two 
dimensions the ferromagnetic model undergoes a phase transition which is 
continuous for $q\leq 4$ and first order otherwise \cite{Baxter}. The Coulomb
gas \cite{Nienhuis} and conformal field theory (CFT) \cite{DF} provided later 
a complete description of the second order phase transition line, which
turned out to correspond to CFT's with central charge $c\leq 1$,
the value $c=1$ corresponding to the end point $q=4$. More recently,
integrable field theory has lead to new results for the scaling 
limit of the off-critical model for $q\leq 4$ \cite{CZ,DC,DBC}.

Concerning the first order transition for $q>4$, several exact lattice 
results -- internal energy \cite{Baxter}, magnetisation \cite{Baxter82}, 
correlation length \cite{BW} -- are known, but progress through field theoretic
methods is generally prevented by the absence of a scaling limit.
When $q$ approaches 4 from above, however, the correlation length at $T_c$
diverges and a continuum description in terms of a {\em massive} quantum field 
theory should be possible. The identification and solution of the quantum
field theory describing the limit $q\rightarrow 4^+$ at $T_c$, as well as
the determination of the universal critical quantities, are the subject of 
this note.

\vspace{.3cm}
The correlation length at criticality of the lattice
model is known to behave as \cite{BW}
\EQ
\xi\sim ae^{\pi^2/\sqrt{q-4}}
\lab{xi}
\EN
as $q\rightarrow 4^+$, where $a$ is proportional to 
the lattice spacing, so that the continuum
limit corresponds to taking the limits $a\rightarrow 0$, $q\rightarrow 4^+$
in such a way that $\xi$ remains finite. The presence of an essential 
singularity rather than a power law divergence in Eq.\,(\ref{xi}) is
characteristic of a
perturbing operator which is only marginally relevant (scaling 
dimension 2) -- except that in the original description of the Potts
model $q$ is a parameter which should be invariant along RG trajectories
and therefore such an interpretation needs to be treated with some care.
In fact we know that \em at \em $q=4$ there is such a marginal field
$\psi$:
it was shown by Nienhuis \em et al. \em \cite{NBRS} to correspond to the
fugacity for vacancies in the lattice model. When $\psi<0$ the
transition is second order, with logarithmic modifications
to scaling arising from the marginal irrelevance of $\psi$, but when 
$\psi>0$, the transition is first order with a correlation length at
the transition which diverges as $\psi\to0^{+}$ like
\EQ
\xi\sim ae^{1/b\psi}
\lab{xi2}
\EN
where $b$ is a constant. 

The main point is that the \em scaling limits \em in which $a\to0$ with
$\xi$ fixed are \em identical \em in these two cases. This may be seen,
for example, from the general structure of the RG equations near $q=4$
and $\psi=0$. Based on the analysis of Nienhuis \em et al. \em
\cite{NBRS}, Cardy, Nauenberg and Scalapino \cite{CNS} argued that these
have, to lowest order, the general form 
\begin{eqnarray}
d\psi/dl&=&b\psi^2+a(q-4)+O\big(\psi^3,(q-4)\psi,(q-4)^2\big)\label{dpsidl}\\
dt/dl&=&\big(y_t+c_t\psi)t+O\big(\psi^2t,(q-4)t,t^2\big)\label{dtdl}\\
dh/dl&=&\big(y_h+c_h\psi)h+O\big(\psi^2h,(q-4)h,th\big)\label{dhdl}
\end{eqnarray}
where $t$ is the deviation from the critical temperature and $h$ is
a symmetry-breaking field. $a$, $b$, $c_t$ and $c_h$ are all constants,
certain combinations of which are universal \cite{CNS}.
An important feature of (\ref{dtdl},\ref{dhdl}) is that, to lowest
nontrivial order, they do not involve $q$. This is because, as may be
seen from the first equation, $q-4$ is effectively $O(\psi^2)$. 
Integrating (\ref{dpsidl}) up to a value $\tilde l$ such that
$\psi(\tilde l)=O(1)$ then gives results for the correlation length
$\xi\sim e^{\tilde l}$ in agreement with (\ref{xi},\ref{xi2}) in the
two cases $q>4$ and $q=4$, $\psi>0$. The various thermodynamic
quantities are then found by integrating the other equations up to
$\tilde l\sim\ln\xi$. To the order stated, the results will be identical
in the two cases, when expressed in terms of $\xi$. Therefore the scaling
limits are identical.

Another way of understanding this is through the mapping of the lattice
Potts model to a height model and thence to a Coulomb gas or sine-Gordon
theory \cite{Nienhuis}. 
{}From this point of view, $q$ is merely a parameter identified
with a certain function of
the coupling constant conventionally called $\beta$. $q=4$
corresponds to $\beta^2=8\pi$ at which point the operator
corresponding to $\psi$ becomes marginal. Within the standard RG
picture of the sine-Gordon model, both $\beta$ and $\psi$ have
non-trivial marginal flows. However the scaling limit, corresponding to
the massive sine-Gordon theory at $\beta^2=8\pi$, is unique, and therefore
describes both the cases $q\to4^{+}$ and $\psi\to0^{+}$ at $q=4$.

Having made this observation, we may take over the results of
Ref.\cite{dilute} in which it was pointed out that the 
scaling limit of the massive $q=4$ theory is integrable, and in which
the scattering theory and form factors were determined.

In our case, the construction of the scattering theory goes as follows.
Along the first order phase transition line, $q$ ordered ground states 
are degenerate with the disordered ground state. The 
field theory describing the scaling limit $q\rightarrow 4^+$ has 4
ordered vacuum states $\Omega_i$, $i=1,\ldots,4$. Invariance under 
colour permutations implies that, in the order parameter space, they lie at 
the vertices of a tetrahedron having the disordered vacuum $\Omega_0$ 
at its center. The elementary excitations of the scattering theory are stable 
kinks $K_{0i}(\th)$, $K_{i0}(\th)$ interpolating between the center of the 
tetrahedron and the $i$-th vertex, and vice versa. We denote by $\th$ the 
rapidity variable parameterising the on-shell momenta as $(p^0,p^1)=
(m\cosh\th,m\sinh\th)$. The mass of the kinks $m\sim\xi^{-1}$ measures the
deviation from the conformal point $q=4$. The space-time trajectory of a kink 
on the plane draws a domain wall separating a coloured phase from the 
disordered one. The space of asymptotic states is made of multi-kink sequences 
in which adjacent 
vacuum indices belonging to different kinks have to coincide. For example, up 
to possible bound states, the lightest excitation interpolating between two
ordered vacua is $K_{i0}(\th_1)K_{0j}(\th_2)$. 

The factorisation of multi-kink processes reduces the scattering problem to
the determination of the two-kink amplitudes. Colour permutation symmetry 
allows only for the four elementary processes depicted in Fig.\,1. The four 
amplitudes can be determined as a solution of the requirements of unitarity
(crucially simplified by the absence of particle production), crossing and 
factorisation. 
The scattering amplitudes of Fig.\,1 are 
given in \cite{dilute} and read
\bea
&& A_0(\th)=\frac{e^{-i\gamma\th}}{2}\,\frac{2i\pi-\th}{i\pi-\th}\,S_0(\th)\,,
\label{a0}\\
&& A_1(\th)=\frac{e^{-i\gamma\th}}{2}\,\frac{\th}{i\pi-\th}\,S_0(\th)\,,\\
&& B_0(\th)=e^{i\gamma\th}\,\frac{i\pi+\th}{i\pi-\th}\,S_0(\th)\,,\\
&& B_1(\th)=e^{i\gamma\th}\,S_0(\th)\,,
\label{b1}
\eea
where $\th$ is the rapidity difference of the two kinks, 
$\gamma=\frac{1}{\pi}\ln 2$, and
\EQ
S_0(\th)=\frac{
\Gamma\left(\frac{1}{2}+\frac{\th}{2i\pi}\right)
\Gamma\left(-\frac{\th}{2i\pi}\right)}{
\Gamma\left(\frac{1}{2}-\frac{\th}{2i\pi}\right)
\Gamma\left(\frac{\th}{2i\pi}\right)}=
-\exp\left\{i\int_0^\infty\frac{dx}{x}\,
\frac{e^{-\frac{x}{2}}}{\cosh\frac{x}{2}}\,
\sin\frac{x\th}{\pi}\right\}\,\,.
\EN
The absence of poles in the physical strip {\mbox Im}$\th\in(0,i\pi)$ 
ensures that there are no bound states and that the four amplitudes above 
completely determine the scattering theory. This $S$-matrix shares 
evident analytic similarities with that of the $SU(2)$-invariant Thirring
model. As a matter of fact, the latter is a realisation of the same perturbed
CFT on a different particle basis (an $SU(2)$ doublet rather than our kinks).
The possibility for a single perturbed CFT to be invariant under different
symmetry groups and to describe different universality classes is discussed, 
for example, in Ref.\,\cite{rsos}.

\vspace{.3cm}
Making contact with the thermodynamics requires the computation of 
correlation functions. In our $S$-matrix framework these are obtained as 
spectral series summing over all multi-kink intermediate states. Neglecting 
terms of order $e^{-4m|x|}$ in this large distance expansion, we will 
approximate the (connected) two-point correlator of a scalar operator 
$\Phi(x)$ as
\EQ
\langle\Omega_0|\Phi(x)\Phi(0)|\Omega_0\rangle\simeq\sum_{i=1}^4
\int_{\th_1>\th_2}\frac{d\th_1}{2\pi}\frac{d\th_2}{2\pi}
|F^{\Phi}_{0i}(\th_1-\th_2)|^2\,e^{-|x|E_2}\,,
\label{approx1}
\EN
in the disordered phase, and
\EQ
\langle\Omega_i|\Phi(x)\Phi(0)|\Omega_i\rangle\simeq
\int_{\th_1>\th_2}\frac{d\th_1}{2\pi}\frac{d\th_2}{2\pi}
|F^{\Phi}_{i0}(\th_1-\th_2)|^2\,e^{-|x|E_2}\,,
\label{approx2}
\EN
in the $i$-th ordered phase. Here $E_2=m(\cosh\th_1+\cosh\th_2)$ is the energy 
of the two-kink asymptotic state and we introduced the two-kink form factors
\bea
&& F^\Phi_{0i}(\th_1-\th_2)=\langle\Omega_0|\Phi(0)|K_{0i}(\th_1)K_{i0}(\th_2)
\rangle\,,\label{ff1}\\
&& F^\Phi_{i0}(\th_1-\th_2)=\langle\Omega_i|\Phi(0)|K_{i0}(\th_1)K_{0i}(\th_2)
\rangle\label{ff2}\,.
\eea

The operators of interest for us are the spin $\sigma_j(x)=
\delta_{s(x),j}-1/q$, and the energy $\varepsilon(x)=\sum_y\delta_{s(x),s(y)}$,
whose scaling dimensions around the $q=4$ fixed point are $X_\sigma=1/8$ and 
$X_\varepsilon=1/2$ \cite{Nienhuis}. 

Some consequences for the physics of the coexisting phases follow immediately
from the structure of the scattering theory. The `true' correlation length
$\xi$ is determined by the large distance decay of the spin-spin 
correlator as $\langle\sigma_j(x)\sigma_j(0)\rangle\sim e^{-|x|/\xi}$. Then
it follows from (\ref{approx1},\ref{approx2}) that 
\EQ
\xi_o=\xi_d=1/2m\,\,
\EN 
(here and below the subscript $o$ ($d$) denotes quantities computed in the
ordered (disordered) phase).
Numerical simulations \cite{JK} and large $q$ expansions \cite{Arisue1}
suggest that the phase independence of $\xi$ holds true for all $q>4$. 

Since the interfacial tension between two coexisting phases is given by the 
total mass of the lightest excitation interpolating between them, we also 
have $\sigma_{od}=m$ and $\sigma_{od}=\sigma_{oo}/2$. The latter result is 
known to hold for all for $q>4$ \cite{BJ}.

The relation of the other interesting thermodynamic quantities (spontaneous 
magnetisation $M$, latent heat $L$, susceptibility $\chi$, specific heat 
$C$, second moment correlation length $\xi_{2nd}$) with the connected 
correlators of $\sigma_j$ and $\var$, and their behaviour as 
$q\rightarrow 4^+$ are
\bea
&& M=\langle\Omega_j|\sigma_j|\Omega_j\rangle\simeq B\,\xi^{-1/8}\,,
\label{param1}\\
&& L=\langle\var\rangle_d-\langle\var\rangle_o
\simeq {\cal L}\,\xi^{-1/2}\,,\\
&& \chi=\int d^2x\,\langle\sigma_j(x)\sigma_j(0)\rangle\simeq\Gamma_{o,d}\,
\xi^{7/4}\,,\\
&& C=\int d^2x\,\langle\var(x)\var(0)\rangle\simeq A_{o,d}\,\xi\,,\nonumber\\
&& \xi_{2nd}^2=\frac{1}{4\chi}\int d^2x\,|x|^2\langle\sigma_j(x)\sigma_j(0)
\rangle\simeq (f_{o,d}\,\xi)^2\,.\label{param2}
\eea
For the ordered case, the two-point correlators of $\sigma_j$ entering $\chi$ 
and $\xi_{2nd}$ are computed on the vacuum $|\Omega_j\rangle$. Since $\var(x)$ 
is odd under the duality transformation exchanging the low- and 
high-temperature phases, we have 
\EQ
\langle\var\rangle_d=-\langle\var\rangle_o\,\,,\hspace{1cm}A_d=A_o\,\,. 
\EN

The critical amplitudes are normalisation dependent but can be combined into a 
series of universal ratios characterising the scaling limit.
We can evaluate the critical amplitudes by integrating the two-particle
approximations (\ref{approx1},\ref{approx2}) of the correlators. What we 
need to know are the two-kink form factors of the operators $\var$ and 
$\sigma_j$. Once again the result is contained in Ref.\,\cite{dilute} and reads
\bea
&& F^\var_{i0,0i}(\th)=\mp\,iL\,
\frac{e^{\pm\frac{\gamma}{2}(\pi+i\th)}}{\th-i\pi}\,F_0(\th)\,\,,\\
&& F^{\sigma_j}_{i0,0i}(\th)=\mp\,M\frac{4\delta_{ij}-1}{6\Upsilon_+(i\pi)}\,
\frac{e^{\pm\frac{\gamma}{2}(\pi+i\th)}}{\cosh\frac{\th}{2}}\,
\Upsilon_\pm(\th)F_0(\th)\,\,,
\eea
with $\Upsilon_-(\th)=\Upsilon_+(\th+2i\pi)$,
\EQ
\Upsilon_+(\th)=\exp\left\{2\int_0^\infty\frac{dx}{x}\,
\frac{e^{-x}}{\sinh 2x}\,\sin^2\left[(2i\pi-\th)\frac{x}{2\pi}\right]
\right\}\,\,,
\EN
\EQ
F_0(\th)=-i\sinh\frac{\th}{2}\,\exp\left\{-\int_0^\infty\frac{dx}{x}\,
\frac{e^{-\frac{x}{2}}}{\cosh\frac{x}{2}}\,
\frac{\sin^2\left[(i\pi-\th)\frac{x}{2\pi}\right]}{\sinh x}\right\}\,\,.
\EN

The results we obtain for the universal amplitude ratios are given
in Table\,1 and compared with those following from the combination of the 
exact \cite{Baxter,Baxter82,BW} and series \cite{Arisue} lattice 
results for the amplitudes.


\begin{center}
\begin{tabular}{|c||c|c|}\hline
             & Field theory & Lattice \\ \hline
$ f_d      $ & $0.6744$ &   $0.673(8)$      \\
$ f_o/f_d  $ & $0.9340$  &  $0.935(5)$     \\ 
$\Gamma_d/\Gamma_o$ & $1.1406$ & $1.19(5)$   \\  
$A_d/{\cal L}^2$ & $0.1047$ & $0.105(3)$      \\
$\Gamma_d/B^2$ & $0.06607$ & $0.0656(15)$ \\ \hline
\end{tabular}
\end{center}
{\bf Table 1.} Universal amplitude ratios for the $q$-state Potts
model at $T=T_c$, $q\rightarrow 4^+$. The field theoretical results are 
obtained within the two-particle approximation.

\vspace{.5cm}
The accuracy exhibited by the two-particle approximation does not come as a 
surprise since it is known as a common feature of this kind of computations 
within integrable field theory. In the present case the accuracy is enhanced
by the low scaling dimensions of the spin and energy operators 
which lead to mild singularities for their correlators and then to a small 
contribution of short distances to the integrals. We estimate that the errors
on our values for the amplitude ratios do not exceed order 0.1\%.

\vspace{.3cm}
The scaling limit we discussed so far corresponds to $q\rightarrow 4^+$. 
At $q=5$, however, the correlation length is still some 2500 times the lattice
spacing and this suggests that our results for $q\rightarrow 4^+$ could still 
provide the basis for an approximate description. 

For a generic value of $q>4$, the model has $q+1$ degenerate ground states at
the transition point  
and the elementary excitations are $2q$ kinks going from the disordered 
vacuum to the $q$ ordered vacua, and vice versa. If the correlation length
is sufficiently large, an approximate scaling should still hold and then
it makes sense to keep for the physical quantities the parameterisations 
(\ref{param1}--\ref{param2}), namely a power of the correlation length
times an amplitude. It is easy to see, however, that in the present case we 
have to allow for a $q$-dependence of the amplitudes. 
Consider in fact the correlator 
\EQ
G_{\alpha j}(x)=\langle\Omega_\alpha|\sigma_j(x)\sigma_j(0)|\Omega_\alpha
\rangle\,,\hspace{.5cm}\alpha=0,i\,;\hspace{.2cm}i,j=1,\ldots,q\,.
\EN
Its two-particle approximation in the disordered phase is 
\EQ
G_{0j}(x)\simeq
\sum_{i=1}^q|F^{\sigma_j}_{0i}|^2 e^{-E_2|x|}=
\frac{q}{q-1}\,|F^{\sigma_j}_{0j}|^2e^{-E_2|x|}\,,
\EN
where integration over momenta is understood. The last equality follows from
colour symmetry and $\sum_j\sigma_j=0$ which imply 
\EQ
\label{rel}
F^{\sigma_j}_{0i,i0}=(q\delta_{ij}-1)/(q-1)F^{\sigma_j}_{0j,j0}. 
\EN
When integrating the correlator to
obtain the amplitude of, say, the susceptibility in the disordered phase, 
the form factors computed at $q=4$ should give a good approximation as long 
as $\xi$ is large. The explicit factor $q/(q-1)$ dictated by the number of 
intermediate states and symmetry, however, has to be taken into account and is 
expected to determine the main deviation from the $q\rightarrow 4^+$ value of 
the amplitude. Following the same reasoning, the susceptibility amplitude
in the ordered phase should be basically constant in $q$  since there is only
one intermediate state. More generally, we are led to expect that the ratios
listed in Table\,2
are approximately constant in $q$ for $\xi$ large 
enough so that the continuum description is accurate.
Their values determined from the results of the large $q$ expansion
and reported in the Table seem to confirm our picture. Our field theory results
for $R_1$ and $R_2$ as $q\rightarrow 4^+$ are 0.855 and 0.0579, respectively.

\begin{center}
\begin{tabular}{|c||c|c|c|}\hline
      q       & $4^+$  & 5       &  10   \\ \hline
$\xi$ (lattice units) &   & 2512.24       &  10.559   \\ 
$\xi_{2nd}^{(d)}/\xi$ & $0.673(8)$ & $0.671(3)$ & $0.6587(1)$     \\
$\xi_{2nd}^{(o)}/\xi_{2nd}^{(d)}$ & $0.935(5)$  &  $0.934(7)$ & $0.9579(2)$ \\ 
$R_1=(q-1)/q\,\chi_d/\chi_o$ & $0.89(4)$ & $0.810(5)$ & $0.80399(1)$   \\  
$R_2=\chi_o/(M\xi)^2$ & $0.0550(6)$ & $0.0589(2)$ & $0.05784(1)$  \\ \hline
\end{tabular}
\end{center}
{\bf Table 2.} Values obtained combining the exact results of 
Refs.\,\cite{Baxter82,BW} and the large $q$ expansions of Ref.\,\cite{Arisue}.

\vspace{.3cm}
Let us conclude this note by considering the `transverse' susceptibility
$\chi_T$ \cite{DBC} obtained integrating $G_{ij}(x)$ rather $G_{jj}(x)$. 
{}From (\ref{rel}),
in the two-particle approximation 
\EQ
\frac{\chi_T}{\chi_o}\simeq\frac{1}{(q-1)^2}\,\,.
\EN
This result is basically a consequence of the nature of the elementary 
excitations and is expected to hold as a good approximation for all $q>4$ at 
$T_c$, as long as $\xi\gg a$. 
We are not aware of lattice results on $\chi_T$ for comparison.

In summary, we have shown that the limit $q\to4^+$ in the Potts model defines
an integrable massive field theory, whose $S$-matrix and form factors may be
computed exactly. The results for integrated correlation functions are in
excellent agreement with lattice-based numerical results. This shows how
methods of continuum field theory are not restricted to the description of
second-order transitions only.

\begin{figure}
\centerline{
\psfig{figure=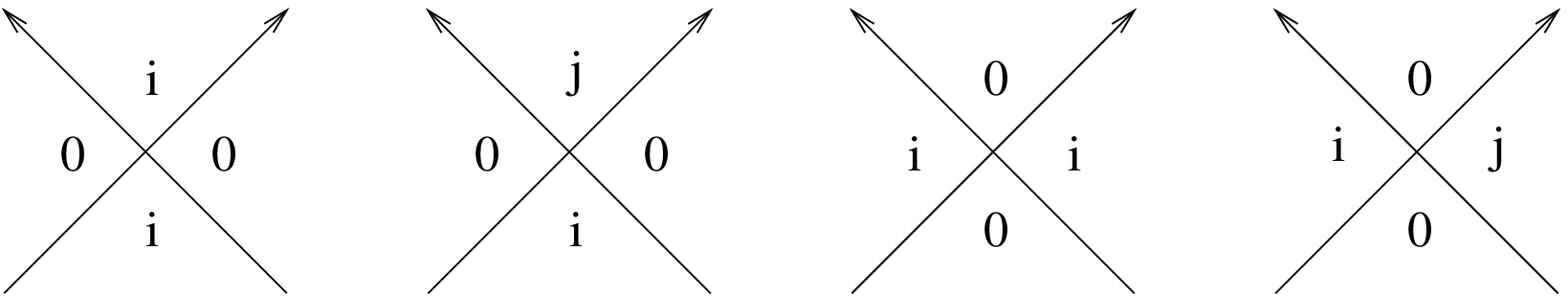}}
{\bf Figure 1.} The two-kink scattering amplitudes $A_0$, $A_1$, $B_0$, 
$B_1$ ($i\neq j$).
\vspace{2cm}
\end{figure}

\end{document}